# Broadband Infrared Spectroscopy of Condensed Phases with Two Intra-pulse Difference-frequency-generation Frequency Combs

Noh Soo Han,[†] JunWoo Kim,[†] Tai Hyun Yoon,[*,†,‡] Minhaeng Cho[*,†,§]

[†]Center for Molecular Spectroscopy and Dynamics, Institute for Basic Science (IBS), Seoul 02841, Republic of Korea

[‡]Department of Physics, Korea University, Seoul 02841, Republic of Korea

[§]Department of Chemistry, Korea University, Seoul 02841, Republic of Korea

*E-mail: mcho@korea.ac.kr (MC) or thyoon@korea.ac.kr (THY)

**ABSTRACT** Mid-infrared (mid-IR) spectroscopy provides a way to study structures and dynamics of complicated molecules in condensed phases. Therefore, developing compact and broadband mid-IR spectrometer has been a long-standing challenge. Here, we show that a highly coherent and broadband mid-IR frequency comb can be generated by using an intra-pulse difference-frequency-generation with a train of pulses from a few-cycle Ti:Sapphire oscillator. By tightly focusing the oscillator output beam into a single-pass fan-out-type periodically-poled lithium niobate crystal and tilting the orientation of the crystal with respect to incident beam, it is shown that mid-IR frequency comb with more than an octave spectral bandwidth from 1550 $cm^{-1}$ (46 THz) to 3650 $cm^{-1}$ (110 THz) and vanishing carrier-envelop offset phase can be generated. Then, using two coherent mid-IR frequency combs, we demonstrate that ultrabroad mid-IR dual frequency comb spectroscopy of both aromatic compounds and amino acids in solutions is experimentally feasible. We thus anticipate that our mid-IR frequency combs could be used to further develop ultrafast and broadband IR spectroscopy of chemically reactive and biological molecules in condensed phases.

## 1. INTRODUCTION

Since the molecular structure is the key to understand its electric and magnetic properties, polarity, chemical reactivity, and biological function, enormous efforts have been put into the development of a variety of spectroscopic techniques enabling to study molecular structure and dynamics in condensed phases. Due to the one-to-one correspondence between molecular structure and mid-IR absorption spectrum, the latter is known as the "molecular fingerprint" of the compound. In the course of a chemical or biological reaction, atomic configurations or molecular geometries of reactants and products change in time, which are clearly reflected in the changes of their transient vibrational spectra in the wavenumber ranges from 1400 to 1800 $cm^{-1}$ and from 2800 to 3500 $cm^{-1}$. Therefore, coherent light sources in the mid-infrared (mid-IR) region from 2 to 20 μm in wavelength have been widely used in a variety of linear and nonlinear IR spectroscopic studies of vibrational transitions and relaxation dynamics of chemical and biological systems.

Although a variety of visible and near-infrared (NIR) coherent radiation sources have been developed and widely applied to time-resolved molecular spectroscopy, fully coherent mid-IR sources with broad spectral bandwidth that covers the entire vibrational spectrum of molecules remain as a long-standing technical challenge mainly due to the lack of broadband gain media. To this end, a few notable coherent mid-IR sources have been developed, which use different approaches such as optical parametric oscillation and amplification[1,2], quantum cascade laser[3,4], and intra-pulse difference frequency generation (DFG)[5-7]. However, these methods have certain limitations in spectral bandwidth, average power, repetition rate, or mutual coherence between successive pulses.

Recently, as an alternative radiation source for molecular spectroscopy in the mid-IR frequency region, optical frequency comb (OFC)[8,9] technologies have been used to develop mid-IR frequency combs for high resolution rovibrational spectroscopy of molecules in condensed phases as well as in the gas phase, where the Fourier-transform (FT) spectra were calibrated directly by the frequency-resolved comb-tooth spacing[6,10,11]. In particular, the dual-frequency comb (DFC) spectroscopy requires two OFCs with slightly different repetition rates, $f_r$ and $f_r + \Delta f_r$. In frequency domain, the spectra of the two OFCs have different comb line spacings. This means that, after a given pair of pulses from the two combs overlaps in time at every $\Delta f_r$, the subsequent pairs of pulses from the two combs have progressively increasing time intervals that are multiples of $\Delta T$ ($\approx \Delta f_r / f_r^2 = f_D / f_r$), where the down-conversion factor of $f_D$ is defined as $\Delta f_r / f_r$. Such automatic time scanning between the two pulse trains from OFCs without any mechanical translational stage[12] has been found to be exceptionally useful for an interferometric detection of free-induction-decay signal in various linear and nonlinear optical spectroscopic studies. To understand its detection process, it should be noted that the interference fringe generated by two OFCs in the time-domain generates a radio frequency (RF) comb with a line spacing of $\Delta f_r$ due to the down-conversion factor of $f_D$ that is one of the experimentally controlled characteristic parameters of a given DFC spectroscopy. Thus obtained down-converted RF signal contains the spectral information of the molecules under investigation.

    For mid-IR DFC spectroscopy in the molecular fingerprint region, various methods to generate mid-IR combs have been reported with varying temporal and spectral coherence, average power, and spectral bandwidth[11]. Mid-IR DFC sources including direct mode-locked lasers[13], optical parametric generation[1,2], quantum cascade lasers[3,4], supercontinuum generation in waveguide[14], Kerr comb generation of microresonator[15], and difference frequency generation[5-7,16] are those examples. The first demonstration of mid-IR DFC spectroscopy is the one reported by

Keilmann *et al.*[7,17], where the mid-IR combs were generated by intra-pulse DFG from nonlinear crystals of gallium selenide (GaSe). Mid-IR combs using such intra-pulse DFG scheme have unique characteristics of the passive stabilization of carrier envelop phase (CEP). Because of the cancelation of CEPs during each DFG process, the generated mid-IR field has zero CEP[18,19] but still the coherence properties of the pump sources are effectively transferred to the two mid-IR frequency combs for DFC spectroscopic measurements. It was also shown that periodically-poled crystals in a quasi-phase matching condition have also been used to enhance the DFC conversion efficiency for mid-IR comb generation[20-23]. Furthermore, oriented-patterned gallium arsenide (OP-GaAs), gallium phosphide (OP-GaP), and periodically-poled lithium niobate (PPLN) have been used to improve the spectral bandwidth and average output power of the mid-IR combs[2,5 6,24].

Mid-IR DFC spectroscopic studies based on different sources were reported over the past years. Especially, mid-IR combs on a chip-scale, which are generated by quantum cascade laser and microresonator, provide the promising potentials for molecular spectroscopy and imaging[25,26]. Although on-chip mid-IR comb spectroscopy has various advantages, such as electrical pumping, large comb line spacing, and high power per each comb line, it has limitation in spectral bandwidth, accuracy of frequency, and flatness of spectral intensity distribution[27-29]. There are a few notable reports demonstrating mid-IR DFC spectroscopy in a broad spectral range[30, 31]. In addition to DFC linear spectroscopy, DFC techniques with coherent anti-Stokes Raman scattering (CARS) in condensed phases have also been demonstrated recently[32-36]. In the DFC-based CARS spectroscopy, the down-converted optical frequency of the CARS signal is measured[32] and the spectral resolution is limited by the spectral width of the light source. This limitation was mitigated by using a fast delay-scan capability of DFC in both time[33]- and frequency[34-36]-domain measurements. In our previous study, we reported transient pump-probe spectroscopy for studying vibronically coupled molecular vibrations of electronic chromophores in solutions, where the spectral resolution is relatively high (< 3 cm$^{-1}$) and the spectral range of vibronically coupled mode is broad (0 – 2700 cm$^{-1}$)[37]. Furthermore, multi-dimensional spectroscopy with multiple combs has been introduced[38-41], which is capable of measuring nonlinear optical spectrum of molecules in condensed phases with both high temporal and spectral resolutions[42], which may be difficult with conventional multi-dimensional spectroscopy methods utilizing mechanical delay lines for multi-dimensional delay time scans.

In the present work, we demonstrate the generation of mid-IR comb with ultrabroad spectral bandwidth (1550 – 3650 cm$^{-1}$) using intra-pulse DFG with a fan-out type PPLN. Two nonlinear crystals were pumped by two stabilized broadband few-cycle NIR Ti:Sapphire oscillators without further amplification. The pump sources have slightly different repetition rates to make the generated mid-IR combs useful for mid-IR DFC spectroscopy of solution samples. Unlike the previous works, to make the broad spectral components (~120 THz) of the pump frequency comb phase-matched simultaneously to generate ultrabroad bandwidth mid-IR combs, the pump beam was tightly focused on each nonlinear crystal, where the angle distribution of input pump beam is wide so that broad spectral components are phase-matched to generate DFG fields. In Sec. 2, we present a detailed description of experimental methods. Results and discussion are given in Sec. 3. The main results will be summarized in Sec. 4 with a few concluding remarks.

## 2. EXPERIMENTAL METHODS

### 2. 1 Chemicals

L-alanine (L-α-aminopropionic acid ≥99.5%, BioUltra, Sigma-Aldrich), deuterium oxide (D$_2$O) (99.9%, Sigma-Aldrich), and toluene (HPLC Plus, ≥99.9%, Sigma-Aldrich) were used in the following experiment. All chemicals were used as received without further purification. Demountable cell (PIKE technologies) with two $CaF_2$ windows and 25.6 μm spacer (FTIR spacers, Harrick Scientific Products Inc.) is used for both FTIR and mid-IR DFC measurement. L-Alanine was dissolved in $D_2O$, where the concentration of the solution was adjusted to be 1 M.

**2. 2 Mid-IR comb generation and mid-IR DFC system**

As shown in Fig. 1(a), to generate mid-IR combs via intra-pulse DFG process, ultra-broadband few-cycle pulse oscillators, denoted as OFC1 and OFC2 (Rainbow 2, Spectra-Physics) were used as pump sources, where two fan-out type periodically-poled lithium niobate (PPLN) (HC photonics) with a thickness of 1 mm and 13.78 – 22.03 μm of poling periods were used. The pump sources with broad spectral bandwidth and the beam waist of 3 mm were tightly focused on the PPLNs with silver-coated parabolic mirror with a focal length of 50.8 mm (MPD129-P01, Thorlabs) and the incident angles were controlled using rotational stages. Thus produced mid-IR combs were collimated with gold-coated parabolic mirrors with the same focal length of 50.8 mm (MPD129-M01, Thorlabs).

The repetition frequencies of the two pump sources were phase-locked to the 18[th] harmonics of 80 MHz and 80 MHz $+\Delta f_r$, respectively, where the 18[th] harmonics were provided by a multichannel RF synthesizer (HS9008A, Holzworth). The detuning frequency $\Delta f_r$ between the two OFCs was adjusted to be 20.56 Hz. Each interferogram is thus sampled with an interval of 12.5 ns ($1/f_r$). The mid-IR combs were combined at ZnSe beam splitter (BSW711, Thorlabs) and then refocused onto the aperture of the MCT detector with a preamplifier (PVI-4TE-6, FIP-1k-1G-F-M4-D, VIGO) by $CaF_2$ lens with a focal length of 250 mm. It should be noted that we observed experimentally the intensity-dependent spectral distortion at high-intensity level so that we attenuated the intensity impinged into the MCT detector until the intensity-dependent spectral distortion disappeared while keeping the signal to noise ratio of the Fourier-transformed spectra reasonably high. The remaining NIR beams were filtered out with a Ge window (Edmund optics) before the mid-IR comb from OFC2 interacts with the target materials. The experimental data of the mid-IR DFC spectroscopy were recorded by a fast digitizer (M3i.4861-EXP, Spectrum) with the sampling rate of 80 MHz, where the signal was referenced by an external clock of 1 GHz from the same RF frequency synthesizer.

**2. 3 Conventional FTIR spectrometer and Mach-Zehnder interferometer**

The absorption spectra of all the samples and PPLN crystals were measured with a conventional FTIR spectrometer (Frontier MIR/FIR spectrometer, PerkinElmer). The obtained experimental data were accumulated 3 times with the spectral resolution of 1 $cm^{-1}$. The same demountable cell was used for both mid-IR DFC and FTIR measurements. To measure the broad spectra of the two NIR pump sources (OFC1 and OFC2) and mid-IR combs, we measured their time-domain interferograms by using a Mach-Zehnder (MZ) interferometer shown in Fig. 2(a). Here, a lock-in amplifier (SR844, Stanford research systems) referenced by RF frequency synthesizer (HS9008A, Holzworth) was used. The mirrors, beam splitters, and detectors used for the MZ interferometric measurements of NIR OFC and mid-IR frequency combs were used differently because of their difference in frequency. To control the relative time delay of the two arms in the MZ interferometer, a delay stage (M-VP-25XL, Newport) and a controller (XPS-Q8 controller, Newport) were used, which could record the absolute position of the delay stage and the analog output signal of lock-in amplifier as schematically described in Fig. 2(a).

## 3. RESULTS AND DISCUSSION

### 3. 1 Ultrabroadband mid-IR comb generation

To generate a train of mid-IR pulses, intra-pulse DFG process was used, where the two fan-out type PPLN crystals with a thickness of 1 mm and poling periods of 13.78 – 22.03 μm were pumped by two ultra-broadband few-cycle pulse oscillators, OFC1 and OFC2, of which properties were described in detail elsewhere[43] (see also Experimental Methods). Briefly, the center wavelengths of the OFC1 and OFC2 are around 800 nm, the pulse duration time is approximately 6 fs, the average output power is about 500 mW, and the spectral bandwidth is approximately 120 THz. The time-domain interferograms and the resulting optical spectra measured by the Mach-Zehnder interferometer are shown in Figs. 2(b) and 2(c), respectively.

In the previous works by other research groups, mid-IR combs were generated via intra-pulse DFG with quasi-phase-matching (QPM) technique to increase the DFG efficiency[2,5,6]. However, the QPM approach has a limitation in the spectral bandwidth due to the fixed phase-matching condition imposed by the constant periodically poling period[6]. Here, the fan-out type QPM crystal and waveguide with continuous polling periods, which could achieve various phase-matching conditions in a single crystal, are used to increase the spectral bandwidth of mid-IR combs and to tune the center wavelength[6,44-46]. Furthermore, we show that the fan-out type PPLN crystal with a tight focusing geometry allows us to overcome the limitation in further increasing the spectral bandwidth[47]. More specifically, by tightly focusing the incident beam into 1 mm PPLN crystal using a parabolic mirror with a focal length of 50.8 mm, we could generate a broad distribution of incident wave vectors so that the entire ultrabroad spectral components in the pump can take part in the effective intra-pulse mid-IR femtosecond pulse generation, mainly due to the relaxed quasi-phase matching condition. To further maximize the spectral bandwidth, we tilted the crystal orientation with respect to the incident beam to make the period of the poles to be chirped. As a result, the spatially chirped crystal periods as well as the radially dispersed spatial wave vector components satisfy very broad phase-matching conditions for the generation of ultrabroad bandwidth mid-IR frequency comb.

Nonetheless, due to the radial distribution of the incident wave vectors and the tilted orientation of the fan-out type PPLN crystals, the output beam in the mid-IR spectral region has a much wider beam divergence angle. In the intra-pulse DFG, any two spectral lines within the pump comb spectrum can act as two incident fields generating a DFG field when they satisfy the DFG phase-matching condition, $\mathbf{k}_{DFG} = \mathbf{k}_1 - \mathbf{k}_2$. The wavevector of the DFG field ($\mathbf{k}_{DFG}$) is not necessarily parallel to the beam propagation direction unless the two incident fields are parallel to each other. Therefore, the DFG field induced by a tightly focused ultra-broadband optical pulse is more dispersed than the incident beam and has a spectral distribution in space. More specifically, the DFG phase-matching condition can be rewritten as a function of the polar angle, $\theta$, with respect to the beam propagation axis as $n_{MIR}\omega_{DFG}\sin\theta_{DFG} = n_1\omega_1\sin\theta_1 - n_2\omega_2\sin\theta_2$, where $n_{MIR}$ ($n_1$ and $n_2$) represents the refractive index of the PPLN at the DFG (pump) frequency of $\omega_{DFG}$ ($\omega_1$ and $\omega_2$)[48]. By assuming that $\theta_1$ and $\theta_2$ are small and that $n_1$ and $n_2$ are similar to each other, the angle of the DFG field, $\theta_{DFG}$, can be simply expressed as $\sin\theta_{DFG} \cong n_1(\omega_1\theta_1 - \omega_2\theta_2)/n_{MIR}\omega_{DFG}$. Thus, the low frequency components in the DFG field should be distributed at the outer region of the beam profile, which is indeed experimentally confirmed. The beam waist of the DFG field was estimated to be about 20 mm – note that the waist of the fundamental beam is 3 mm. Therefore, from this experimental observation and further confirmation with the DFC spectra, it becomes clear that the spatial distribution of ultrabroad bandwidth of the incident beam and the tight focusing at the tilted fan-out PPLN crystal are responsible for the generation of ultra-broadband mid-IR frequency comb. Without these geometries, one could only obtain a mid-IR comb with a relatively narrow spectral bandwidth because of the fixed quasi-phase matching condition of the PPLN crystal.

**3. 2 Ultrabroadband mid-IR DFC spectroscopy of condensed phases**

As a proof-of-principle experiment, the generated ultra-broadband mid-IR comb sources are used to demonstrate a mid-IR DFC linear spectroscopy of solution samples containing either small aromatic compounds or amino acids. Fig. 1(a) shows the schematic diagram of our mid-IR DFC spectroscopy setup. For mid-IR DFC linear absorption spectroscopy, the repetition rates of OFC1 and OFC2 are 80 MHz and 80 MHz $+\Delta f_r$, respectively, and the detuning frequency $\Delta f_r$ is 20.56 Hz. Therefore, the period of the interferograms is about 50 milliseconds. The data acquisition was achieved with a fast digitizer whose sampling rate is synchronized to $f_r$ = 80 MHz as well. In the case of mid-IR sources generated by the intra-pulse DFG processes, it is well-known that the carrier-envelop-offset frequency $f_{ceo}$ has in principle to be zero[49]. In Figure 1(b), the time-domain reference interferogram of our mid-IR DFC for a blank cell is depicted. Note that the effective time of the raw interferogram in Fig. 1(b) was obtained from the laboratory time multiplied by the down-conversion factor of $f_D = \Delta f_r / f_r = 2.57 \times 10^{-7}$.

Figure 1(c) shows the ultrabroad spectral feature of the generated mid-IR combs covering from 1550 cm$^{-1}$ to 3500 cm$^{-1}$. The output power of the mid-IR comb is weaker than those of the previously reported ones using DFG processes[5,6]. Although we did not use high-power amplified pump sources with low repetition rates, the generated mid-IR combs are sufficiently strong for detecting the down-converted RF signal from DFC measurement. Considering the voltage responsivity of the pre-amplified detector, $2.3 \times 10^4$ V/W, and the detected voltage of about 1 V, which was measured using a lock-in amplifier, we estimate the output power of the generated mid-IR combs to be approximately 40 μW. Here, it should be noted that because of the strong absorption of the PPLN crystals at around 1750 cm$^{-1}$ as shown in Fig. 3, the spectra of the two mid-IR combs show a broad dip at 1750 cm$^{-1}$. Furthermore, the PPNL crystal strongly absorbs IR beams at wavenumbers below 1500 cm$^{-1}$. This indicates that the absorptive characteristics of the nonlinear crystal limits the spectral bandwidth of the generated mid-IR combs in this frequency domain. To avoid the self-attenuation of generated IR beams at around 1750 cm$^{-1}$ by the crystal, one might have to use different nonlinear crystals such as orientation-patterned GaAs for subharmonic generation[2].

In Fig. 1(c), many sharp absorption lines in the ranges from 1500 cm$^{-1}$ to 2000 cm$^{-1}$ and from 3000 cm$^{-1}$ to 3750 cm$^{-1}$ appear. However, they originate from the rovibrational transitions of water vapor. In addition, the two dips (absorptions) at around 2350 cm$^{-1}$ are the rovibrational band of the asymmetric stretch mode of $CO_2$ in the air. To confirm these assignments, we directly compare the spectral positions in our mid-IR DFC spectrum with those of $H_2O$ and $CO_2$ gases that can be found in the HITRAN database (see the gas phase $H_2O$ and $CO_2$ IR spectra shown in the lower panel of Fig. 1(c) as well as those in Fig. 4)[50].

The frequency resolution of FT spectrum is mainly determined by the sampling frequency ($f_s$) and the total number of data ($N$) of the interferograms. In our experiments, as shown in Fig. 1(b), the DFC interferograms were measured with $f_s$ = 80 MHz and $N$ = 8200 points, respectively, which suggests that the frequency interval of the FT spectrum ($f_s/2N$) is approximately 5 kHz. Then, by considering the down-conversion factor $f_D$, the frequency resolution in our mid-IR DFC spectrum is quantitatively estimated to be approximately 20 GHz (0.7 cm$^{-1}$). Indeed, within the present data acquisition condition, the Lorentzian linewidth of the absorption peak at 1576 cm$^{-1}$ (Fig. 4(e)) is about 2 cm$^{-1}$, which is consistent with our estimated frequency resolution, even though the spectral resolution from comb-tooth resolved spectrum could not be provided here. In the case of the gas phase spectroscopy, high frequency resolution of a given spectrometer for measuring narrow spectral lines is important. However, for molecular spectroscopy of condensed phases, such high frequency resolution capability is not always needed. Usually, vibrational dephasing times of molecular vibrations in solutions are on the order of picoseconds. Therefore, the frequency resolution of 0.7 cm$^{-1}$ in our mid-IR DFC system is sufficiently high enough to resolve various absorption bands of polyatomic molecules in condensed phases. The spectral resolution of mid-IR DFC spectrum could be improved in our system by changing the data acquisition parameters for high resolution mode, where an optical trigger signal should be used to synchronize the DFC interferogram with that of the measurement sequence very precisely, comb-tooth resolved detection[6,10,31], and simultaneous detection of reference signals for the correct phase normalization[6,10,31].

Here, it should be noted that our mid-IR frequency combs are generated by using near-IR frequency comb laser oscillators. Therefore, the long-term stability in intensity is very high compared to optical amplifier based laser. Indeed, the fact that the low-intensity fluctuation of our measured DFC spectra of blank cell and liquid toluene in Fig. 5 supports this. Over the measurement time of 1 s, a series of absorption spectra of liquid toluene are plotted in Fig. 5(a) as a function of laboratory time. Each absorption spectrum is obtained by taking the ratio of the two power spectra of blank cell and liquid toluene, where each power spectrum is the Fourier-transformed one from a single DFC interferogram. Their integrated intensities show almost constant in measurement time (Fig. 5(b)), supporting that the mid-IR comb intensities are quite stable. In addition, the extraordinary stability of down-converted frequency originates from the passive stabilizations of both CEP and repetition frequency that are synchronized to the standard frequency from the GPS disciplined atomic clock. The high stability of the repetition frequencies of our mid-IR combs is confirmed with the experimentally measured Allan deviations of the repetition rates of the two mid-IR combs (Fig. 6).

In Figs. 7(a) and 7(b), we show the raw interferograms and their FT spectra, respectively, of a blank cell as the reference and liquid toluene representing aromatic compounds with a single alkyl group. In Fig. 7(b), solid lines are averaged FT spectra (20 times) of blank cell and liquid toluene, while dotted lines with open circles correspond to those obtained from each single interferogram. The IR absorption spectrum of liquid toluene is obtained by taking the ratio of the two power spectra of the reference cell and the toluene. For comparison, the absorption spectrum of toluene at the same condition was measured by using a conventional FTIR spectrometer, which is the black line in Fig. 7(c). Interestingly, the absorption spectrum from a single interferogram exhibits all the characteristic absorption bands of liquid toluene across a broad spectral region. As shown in Fig. 7(a), the single interferogram with spans of ~102 μs is Fourier-transformed to obtain the FT spectrum (Fig. 7(b)), which is consistent with passive mutual coherence time of mid-IR combs in previous reports[34,51]. This clearly demonstrates that the IR spectrum of condensed phase sample could be measured from just a single interferogram without using additional detection of reference signals for phase correction. From the millisecond-level data acquisition stability of our mid-IR DFC spectroscopy, we expect that this mid-IR DFC linear spectroscopic method, which have the period of interferogram of 50 millisecond, could even be of use for monitoring the time evolution of absorption spectra of various chemically or biologically reactive samples in condensed phases for the long term measurement from seconds to hours[30]. Furthermore, the period of interferogram can be reduced down to 1 ms by increasing the repetition frequency by a factor of 3.5, for example, by employing commercially available mid-IR combs with the repetition frequency of 250 MHz, so that the fast scan with millisecond time-scale could be achieved. However, the figure of merit of our mid-IR DFC system, which is obtained by the product of the average signal-to-noise ratio per unit square root of measurement time and the number of elements[52], is not high enough compared to those of previous studies[6,10,51]. In the present work, the signal to noise ratio at single wavelength can be estimated by the standard deviation of the distinct peaks in the absorption spectra, even though it is not exactly the value of the figure-of-merit. In the case of liquid toluene, the standard deviations of the noises of DFC spectrum from a single interferogram and of that from the average (20 times) interferogram (Fig. 7(c)) in the wavenumber range from 2650 cm$^{-1}$ to 3130 cm$^{-1}$ are estimated to be about 0.03 and 0.006, respectively. Considering the intensities of 0.94 at 3027 cm$^{-1}$ and 0.04 at 2733 cm$^{-1}$ of vibrational modes with high and low intensities in Fig. 4(c), the signal to noise ratios are to be 31.3 and 1.3 for DFC (single) and 156.6 and 6.6 for DFC (average). Furthermore, to estimate the figure of merit of our system, the average signal to noise ratio of Fig. 7(c) in the wavenumber range from 2650 cm$^{-1}$ to 3130 cm$^{-1}$ are derived from the reciprocal of the coefficient of variation, which is the ratio of mean value of signal of 0.16 of DFC (single) and 0.14 of DFC (average) to the standard deviation of the noise. Concerning the standard deviation of the noise, the measurement times of 100 μs versus 1 s, and the spectral elements of 740, the figure of merits of DFC (single) and DFC (average) are estimated to be $3.9 \times 10^5$ s$^{-1/2}$ and $1.7 \times 10^4$ s$^{-1/2}$, respectively. Quite often, the frequency shift and lineshape change upon increasing temperature, concentration change, molecular conformational transition etc. are significantly more important experimental observables in solution-phase vibrational spectroscopy. Here, the frequencies of our mid-IR DFC spectra are corrected by using the sharp absorption line of the symmetric bending mode of H$_2$O vapor at 1654 cm$^{-1}$. The reason why we need to calibrate the frequency of our spectrum is because we use self-triggering measurement of the interferograms, where no additional reference or triggering scheme for data collection are used. We note that, in our experimental condition with a low duty cycle of only $2\times10^{-3}$, the unused optical power during the dark period may induce an unwanted photo-induced thermal effect. To solve this problem, one could use the technique of gated sampling, i.e., using a synchronized optical shutter, to record the time-dependent change of chemical reactions without any thermal degradation[37].

To obtain the DFC spectra of liquid toluene, the baselines of the mid-IR DFC spectra had to be corrected by using the procedure described in detail in Fig. 8. As can be seen in Fig. 7(c), the agreements between our DFC spectra and the FTIR spectrum are quantitative. The CH stretching modes of phenyl group in toluene (3087 cm$^{-1}$, 3062 cm$^{-1}$, and 3028 cm$^{-1}$) and the CH stretching modes of the alkyl group (2949 cm$^{-1}$, 2920 cm$^{-1}$, 2873 cm$^{-1}$) are clearly visible in our mid-IR DFC spectrum (see Refs.[53, 54] for mode assignments). Furthermore, in the low frequency region, even weak overtones of aromatic vibrational modes (1943 cm$^{-1}$, 1858 cm$^{-1}$, 1803 cm$^{-1}$) and CC stretching mode (1605 cm$^{-1}$) can be found. Although an absorption band at 1730 cm$^{-1}$ cannot be measured due to the self-absorption of the difference frequency-generated IR beams by PPLN crystal itself, all the other fingerprint modes of toluene in the high frequency region are clearly measured even by taking into consideration one interferogram (Fig. 7(c)). This demonstrates the sensitivity of our mid-IR DFC spectrometer.

To show potential use of our mid-IR DFC spectrometer for biological systems, we considered zwitterionic L-alanine dissolved in D$_2$O. Because the L-alanine zwitterion has an ammonium cation and a carboxylate anion in neutral solution, their molecular structures strongly depend on pH. To measure the CH stretch IR spectra of L-alanine, we chose $D_2O$ as a solvent that has a broad OD absorption from 2250 $cm^{-1}$ to 2750 $cm^{-1}$. Figure 9(a) shows the interferograms of a blank cell, $D_2O$, and L-alanine dissolved in $D_2O$. Their FT spectra are shown in Fig. 9(b), where the solid lines are the average spectra of twenty FT spectra and the dotted lines with empty circles are those from a single interferogram. Both the absorption spectrum obtained from a single interferogram and the average spectrum over twenty FT spectra show fingerprint bands of L-alanine in deuterated water. Especially, the weak peaks at 2990 $cm^{-1}$ and 2951 $cm^{-1}$ are associated with asymmetric $CH_3$ stretching vibrations and the sharp peak at 1613 $cm^{-1}$ is with asymmetric $CO_2^-$ stretching vibration[55-57]. The broad peak at 3400 $cm^{-1}$ corresponds to the OH stretch band of HDO in $D_2O$, where the $NH_3^+$ in the L-alanine is deuterated in $D_2O$ solution and a small amount of HDO is thus produced. Since the FT spectra even from a single interferogram shows all the characteristic vibrational modes of L-alanine, it is believed that this is another evidence showing that our mid-IR DFC spectroscopy could be of use to study polypeptides in water.

## 4. CONCLUSIONS

In summary, here we have demonstrated an ultra-broadband mid-IR DFC spectroscopy by using two phase-stabilized mid-IR combs with more than one-octave spectral bandwidth, which were produced by employing the intra-pulse DFG technique with fan-out PPLN crystals. Our mid-IR combs provide the broadest spectral coverage so far. Using two mid-IR combs, we carried out mid-IR DFC spectroscopy experiments for liquid toluene and L-alanine in $D_2O$ solution to show that our mid-IR combs are indeed broadband coherent radiation sources in the mid-IR frequency region. As shown here, the mid-IR DFC spectroscopy has advantages of a fast scan rate and long-term phase stability. We thus believe that the present mid-IR frequency comb sources combined with various DFC techniques will be useful to develop novel nonlinear IR spectroscopic or label-free vibrational imaging techniques in the near future.

AUTHOR INFORMATION

**Corresponding Author**

*E-mail: thyoon@korea.ac.kr, mcho@korea.ac.kr

**Notes**

The authors declare no competing financial interests.

ACKNOWLEDGMENT

This work was supported by the Institute for Basic Science (IBS) in the Republic of Korea with Grant No. IBS-R023-D1.

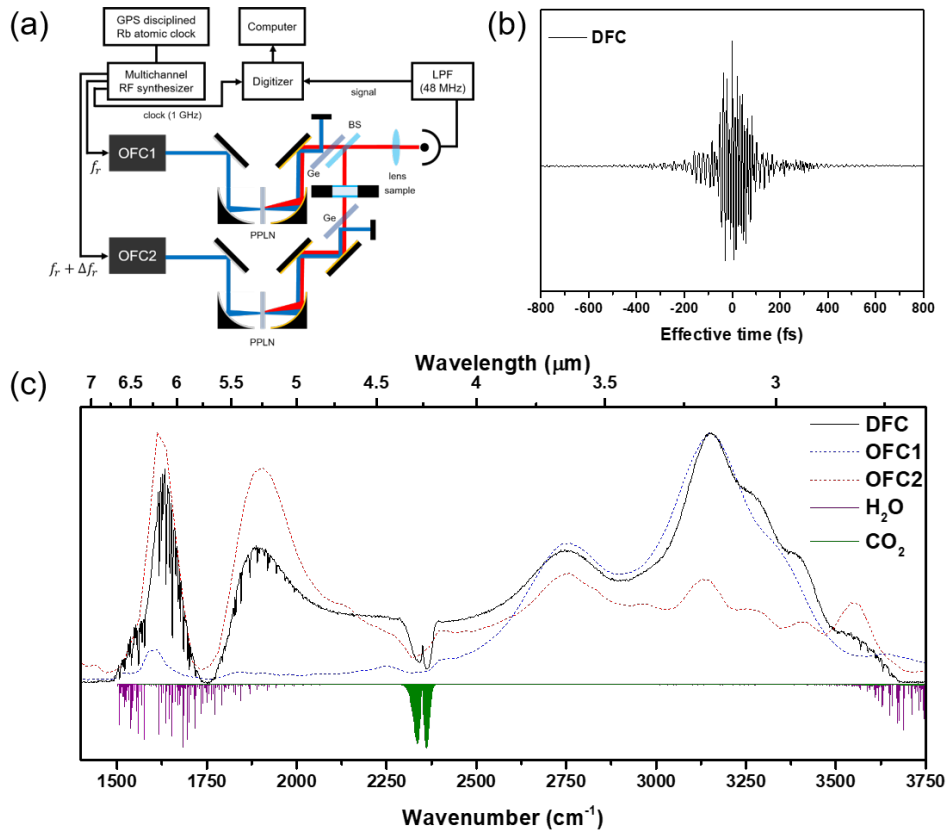

**Figure 1. (a)** Schematic diagram of our mid-IR dual frequency comb spectroscopy setup. Blue and red lines correspond to NIR OFC and mid-IR frequency comb, respectively. The two pump OFCs, denoted as OFC1 and OFC2, have repetition frequencies of $f_r$ and $f_r + \Delta f_r$, respectively. Each NIR OFC pump is focused on fan-out type PPLN crystal using a silver parabolic mirror and then the produced mid-IR comb is defocused using a gold parabolic mirror. Mid-IR comb generated by the OFC2 passes through the solution sample and the transmitted mid-IR light is combined with that by the OFC1 at ZnSe beam splitter. The RF signal resulting from the interference between the two is detected by an MCT detector with a preamplifier. The DFC spectroscopic signal is recorded by the digitizer, where the reference frequency of 1 GHz provided by a multichannel RF synthesizer is used. **(b)** Mid-IR dual frequency comb interferogram without a sample is shown. **(c)** Fourier transformed spectra of various interferograms. Here, the measurement time is approximately 1 s. The dotted lines with blue and red colors are the Fourier transformed spectra of the mid-IR frequency combs generated by the DFG with OFC1 and OFC2, respectively. They were measured by using a Mach-Zehnder interferometer. In the lower panel of this figure, the absorption lines of $H_2O$ and $CO_2$ gas in the HITRAN database are shown as purple and olive colors, respectively, for the sake of direct comparisons.

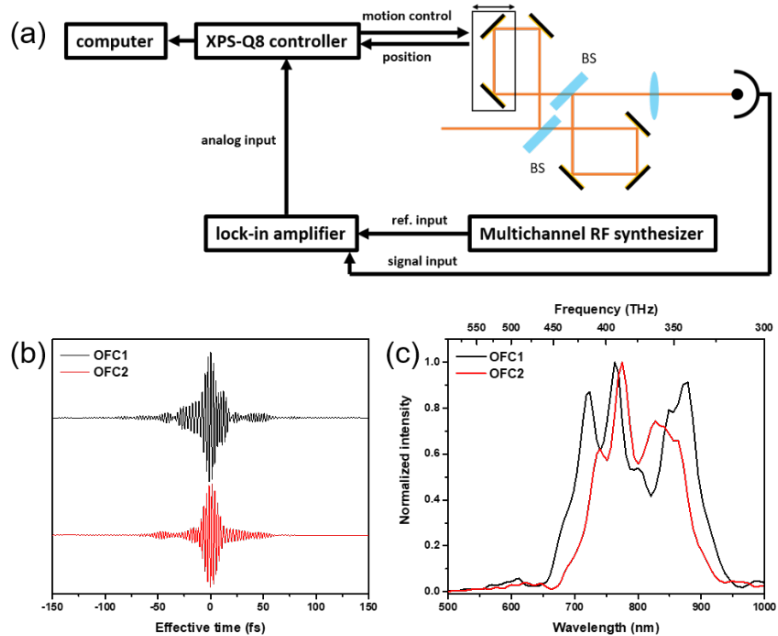

**Figure 2.** (a) Schematic diagram of the Mach-Zehnder interferometer setup used to measure the optical spectra of NIR pump sources and the IR spectra of mid-IR combs. Incident beam is separated by the first beam splitter into two and then later they are combined by the second beam splitter. The time delay between two beams is controlled with a motorized translational stage. The recombined beam is focused on the detector. The signal is sent to a lock-in amplifier to improve the signal to noise ratio. The analog signal from the lock-in amplifier and the position data of the delay stage are recorded by the XPS-Q8 controller, simultaneously. (b) The time-domain interferogram of OFC1 and OFC2 are shown. (c) The average of two spectra of OFC1 and OFC2 are shown.

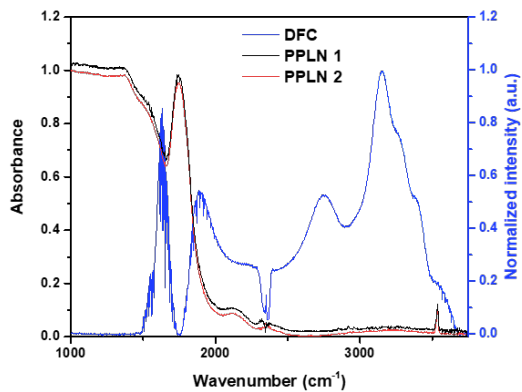

**Figure 3.** The power spectrum (blue line) of the mid-IR DFC spectroscopy data without the sample is shown, which corresponds to the spectrum in Fig. 1(c). The black and red lines are the absorbance spectra of our two PPLN crystals directly measured with the FT-IR spectrometer.

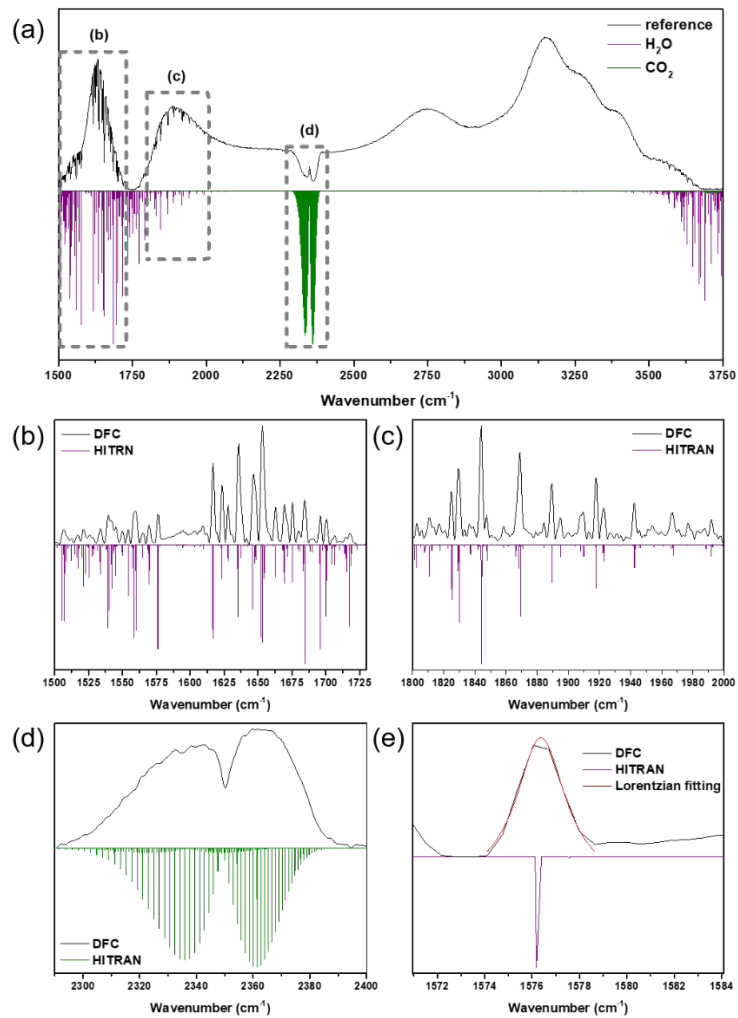

**Figure 4.** (a) The black line in the upper panel shows the power spectrum of mid-IR DFC data for a blank cell. (b)-(d) For the line assignments and comparisons, the absorption spectra of $H_2O$ (purple) and $CO_2$ (olive) in the gas phase that are taken from HITRN database are plotted here. The water bending and OH stretch rovibrational bands in our mid-IR DFC spectrum are in agreement with those in the HITRAN spectrum. In addition, the P and R branches of the asymmetric $CO_2$ stretch rovibrational spectrum are observed in our mid-IR DFC spectrum too. One of the distinctive absorption peaks of $H_2O$ vapor shown in panel (e) is fitted with a Lorentzian function and its full-width-at-half-maximum is found to be approximately 2 cm$^{-1}$.

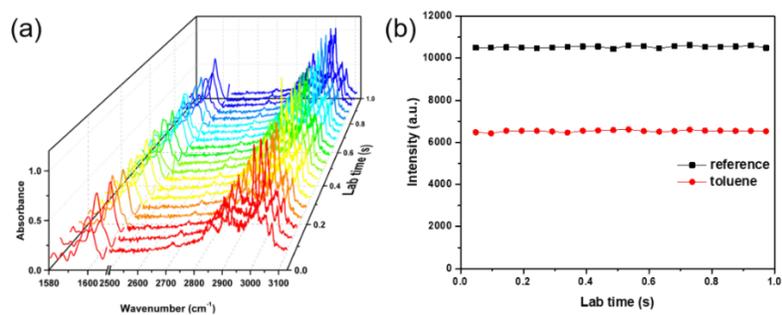

**Figure 5.** **(a)** A series of absorption spectra of liquid toluene are plotted, which were obtained from 20 time-domain interferograms measured over 1 s. **(b)** the integrated intensity of each mid-IR DFC spectrum obtained from the interferograms is plotted with respect to the measurement lab time. They remain constant, which suggests that the stability of our mid-IR comb system is high enough to monitor the time evolution of the IR spectrum for slow (> ms) chemically reactive systems with our mid-IR DFC spectrometer.

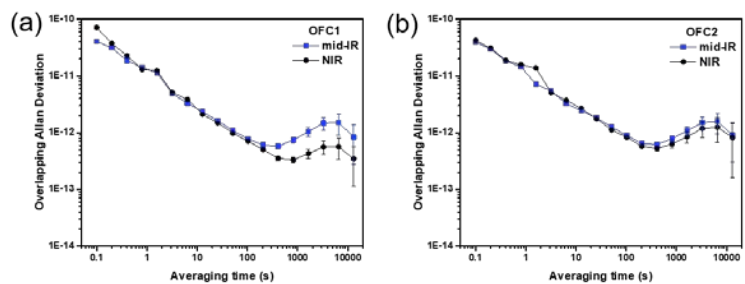

**Figure 6.** The time series of repetition rates of NIR pump sources and mid-IR combs are recorded by a frequency counter. These two figures in (a) and (b) show the overlapping Allan deviation of NIR pump sources and mid-IR combs from OFC1 and OFC2 over the measurement time of 20 hours. Such long-term stability originates from the active stabilization of the repetition rates of OFC1 and OFC2. Intrinsic cancellation of carrier-envelop-offset phases (CEP) during each intra-pulse DFC helps the stability of our mid-IR DFC spectroscopy system.

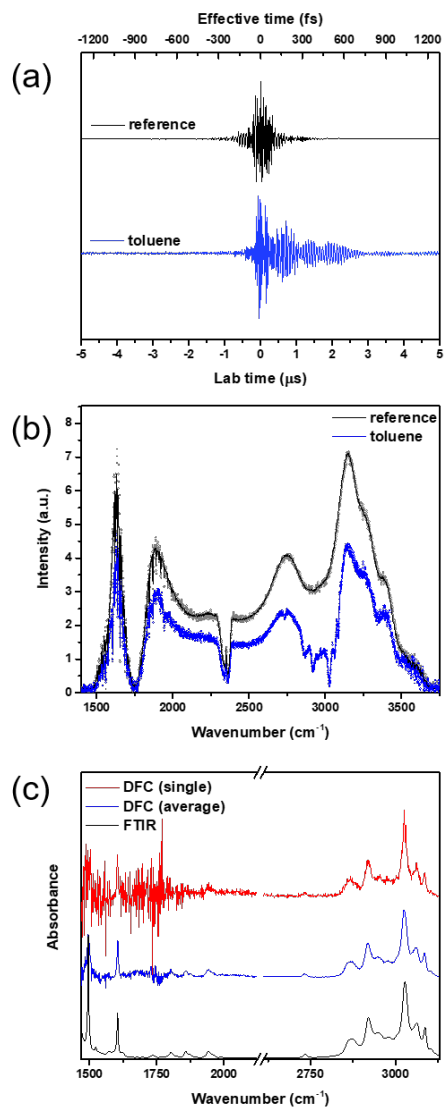

**Figure 7. (a)** Mid-IR DFC interferograms of the blank cell and liquid toluene. **(b)** The corresponding Fourier-transformed spectra are shown. Here, the demountable liquid cell thickness is 25.6 μm. To obtain the FT spectrum, a single interferogram with spans of ~102 μs centered at the center burst (8200 points) is Fourier-transformed. In **(b)**, the solid lines are the averaged one over 20 spectra. The dotted lines with empty circles in gray and blue are the FT spectra recovered from a single interferogram. **(c)** The normalized absorption spectra obtained with a conventional FT-IR spectrometer (solid black line) and our mid-IR DFC spectra (blue and red) are shown. All the absorption spectra are normalized for the sake of comparison. The baselines of mid-IR DFC spectra were corrected by using the procedure presented in Fig. 8.

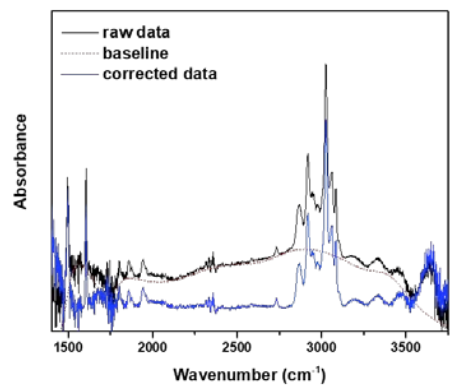

**Figure 8.** The raw spectrum (black) before baseline correction and the corrected spectrum (blue) are shown. The red dotted line is the baseline that should be subtracted out from the raw spectrum. The baseline is created by the spline interpolation based on anchor points that can be found by examining the 2$^{nd}$ derivative spectrum.

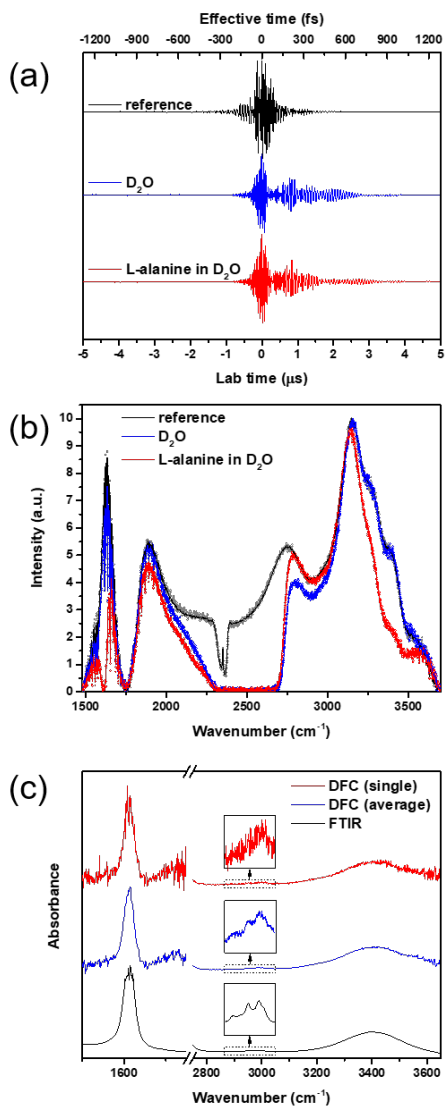

**Figure 9. (a)** Mid-IR DFC interferograms of the blank cell, pure liquid $D_2O$, and L-alanine in $D_2O$. **(b)** The corresponding Fourier-transformed spectra are shown. The solid lines are the average of 20 spectra, whereas the dotted lines with empty circles of gray, blue, and red colors are the FT spectra obtained from a single interferogram. **(c)** The normalized absorption spectra measured with conventional FT-IR spectrometer and our mid-IR DFC spectroscopic technique are shown. Again, the mid-IR DFC (single) and DFC (average) spectra were corrected by subtracting the background baselines in Fig. 8.